\documentclass[aps,prl,twocolumn,showpacs,preprintnumbers,amsmath,amssymb]{revtex4}
\usepackage{epsfig}
\usepackage{epstopdf}
\usepackage{amsmath}
\usepackage{lipsum}
\usepackage{xcolor}

\begin{document}
	\title{Truncated L\'{e}vy Walks and Superdiffusion\\in Boltzmann-Gibbs Equilibrium of the Hamiltonian Mean-Field Model}
	\author{ Piotr Fronczak, Agata Fronczak, Anna Chmiel, Julian Sienkiewicz}
	\affiliation{Faculty of Physics, Warsaw University of Technology,
		Koszykowa 75, PL-00-662 Warsaw, Poland}
	\date{\today}
	
\begin{abstract}
The Hamiltonian Mean-Field (HMF) model belongs to a broad class of statistical physics models with non-additive Hamiltonians that reveal many non-trivial properties, such as non-equivalence of statistical ensembles, ergodicity breaking, and negative specific heat. With this paper, we add to this set another intriguing feature, which is that of super-diffusive equilibrium dynamics. Using molecular dynamics techniques, we compare the diffusive properties of the HMF model in the quasi-stationary metastable state (QSS) and in the Boltzmann-Gibbs (BG) regime. In contrast to the current state of knowledge, we show that L\'evy walks underlying super-diffusion in QSS do not disappear when the system settles in the thermodynamic equilibrium. We demonstrate that it is extremely difficult to distinguish QSS from the BG regime, by only examining the statistics of L\'evy walks in HMF particle trajectories. We construct a simple stochastic model based on the truncated L\'{e}vy walks with rests that quantitatively resembles diffusion behavior observed in both stages of the HMF dynamics.
\end{abstract} \maketitle

In recent years, long-range interacting systems have become a fascinating research area \cite{bookCampa,Campa2009}. This has occurred because of some unusual properties resulting from the non-additivity of their energy (see, e.g., ~\cite{bookCampa} and references therein). The most intriguing features of these non-additive systems are \cite{1999LyndenBell,1999PRETorcini,Latora_2000,2000JStatPhysEllis,2001PRLBarre,Pluchino_2004,2004PhysAEllis,2005PRLMukamel,2006PRLaBaldovin,2007PhysACampa,2010PRLKastner,2013JStatMechMori,2017PREHovhannisyan,2018PREBaldovin}: negative specific heat, breakdown of ergodicity, nonequivalence of statistical ensembles, and slow relaxation processes, which are observed as quasi-stationary states (QSSs), whose lifetimes diverge as the system size increases. The Hamiltonian Mean-Field (HMF) model \cite{1995PREAntoni, 1998PRLLatora, 1999PRLLatora}, which we are dealing with in this Letter, describes $N$ classical particles moving on the unit circle and interacting through an infinite range potential and is one of the best-known models that fall into this class of systems. 
	
In what follows, we report on some non-trivial properties of the HMF model which, despite numerous studies, have escaped the attention of researchers. To be more specific, we confront the dynamical properties of the model in the metastable QSS with its properties in the Boltzmann-Gibbs (BG) regime. Contrary to what has been claimed so far \cite{1999PRLLatora, 2000PhysALatora}, we show that the super-diffusive dynamics of this model, which was thought to be a distinctive feature of its QSS, does not change into normal diffusion, when the system settles into equilibrium. In this regard, our research confirms previous observations, according to which, in the model studied, the long-standing metastable QSS is microscopically very similar to the BG equilibrium \cite{2006PRLbBaldovin, 2007PhysATsallis}. We find it remarkable that, although from the macroscopic point of view, by using the concept of kinetic temperature, it is easy to determine whether the system is (or not) in QSS, from the microscopic point of view, it is a very difficult task. At least, the issue cannot be solved when only diffusive properties of the model, resulting from single-particle trajectories, are examined, as was claimed in Ref.~\cite{1999PRLLatora}.
	
In Ref.~\cite{1999PRLLatora}, the authors studied the molecular dynamics of the HMF model in the quasi-stationary state, when the system is started from out-of-equilibrium initial conditions. They found that, in QSS, the motion of particles is super-diffusive, that is, time dependence of their average square displacement is given by $\sigma^2_\theta(t)\sim t^\alpha$ with $\alpha>1$. They noticed, that after a cross-over time, $t_c$, the diffusion becomes normal, with $\alpha=1$. Moreover, the cross-over time was found to coincide with the time, $t_r$, the system needs to settle in the equilibrium state, which, in turn, is usually treated as being the same as the time when the kinetic temperature of the system agrees with that analytically obtained within the canonical treatment \cite{1995PREAntoni}. In Ref.~\cite{1999PRLLatora}, it was also argued that the super-diffusive motion of HMF particles can be understood in terms of L\'{e}vy walks \cite{Zaburdaev_2014}, according to the following scenario: Each particle performs a quasi-regular trapped motion around a cluster formed by other particles, or undergoes free walks far from it. After a cross-over time, L\'{e}vy walks disappear and super-diffusion turns into the normal one. It happens when the system approaches BG equilibrium. 
	
We show that although the general picture of the HMF dynamics in QSS, which was given in Ref.~\cite{1999PRLLatora}, is true, some findings of~\cite{1999PRLLatora}, which also have been numerously acknowledged by other authors \cite{Latora_2000,Pluchino_2004,Yamaguchi_2004, Levin_2014}, are clearly wrong.  In particular, L\'{e}vy walks are not only observed in QSS, but they are also frequent in BG equilibrium. Regardless of whether the system is in QSS or not, trajectories of individual particles consist of intertwining periods in which particles rest and those in which they move with a constant velocity. Interestingly, the probability distributions characterizing the duration of such 'rests' and 'walks' for the metastable state are almost identical to those for the thermodynamic equilibrium. In both cases, the distributions are well fitted by power-laws with an exponential cut-off. We demonstrate that it is this cut-off that causes the exponent $\alpha$ in $\sigma^2_\theta(t)$ to change its value from $\alpha>1$ to $\alpha=1$. In fact, this behavior of $\alpha$ has nothing to do with the transition of the system from the out-of-equilibrium QSS to BG equilibrium but is also visible when the system is already BG regime for a long time. Surprisingly, however, the transition QSS $\rightarrow$ BG manifests itself as the plateau $\alpha\simeq 1.5$ which is observed for $t<t_r$. Consequently, it is not true that $t_c$ coincides with $t_r$. 

We prove the above findings in two ways. First, by proposing a simple stochastic model based on the truncated L\'{e}vy walks with rests, which quantitatively resembles the diffusion behavior observed in the model. Secondly, by performing a kind of renormalization of particle positions, which allows to study the dispersion of their trajectories starting at various stages of the HMF model dynamics, without interfering with its temporal evolution. In the following we recall the HMF model, and then we demonstrate our results.

The HMF model describes a system of $N$ interacting classical particles (or rotors) characterized by the angles $\theta_i$ and the conjugate momenta $p_i$. Its Hamiltonian is given by
\begin{equation}
{\cal H}(\theta,p)=K+V,
\label{hamiltonian}
\end{equation}
where 
\begin{equation}
K=\frac{1}{2}\sum_{i=1}^N p_i^2,\;\;\;\;\;V=\frac{1}{2N}\sum_{i,j = 1}^N \left[1-\cos(\theta_i - \theta_j)\right]
\end{equation}
are the kinetic and the potential energies, respectively. If we assign to each particle a spin vector $\textbf{m}_i=\left[\cos(\theta_i), \sin(\theta_i)\right]$, then the modulus of the magnetization $\textbf{M}=\sum_{i=1}^N \textbf{m}_i=[M_x,M_y]$ will serve as the order parameter of the model, in a close analogy to the fully-coupled XY model. The HMF model is solvable in the canonical ensemble \cite{1995PREAntoni} and exhibits the second-order phase transition from para- to ferromagnetic phase at the critical energy $U_c=E_c/N=0.75$, which corresponds to the critical temperature $T_c=0.5$. Rewriting the potential energy as $V=N[1-(M_x^2+M_y^2)]/2$, one obtains the equations of motion for all the particles:
\begin{equation}
\frac{d\theta_i}{dt}=p_i,\;\;\;\;\; \frac{dp_i}{dt}=-\sin(\theta_i)M_x+\cos(\theta_i)M_y.
\label{rownruchu}
\end{equation}
The molecular dynamics of the HMF model can then be examined by a numerical integration of the above equations. 

In our numerical studies, the system size is always $N=2048$. Also, following the other authors \cite{Cirto_2014,Cirto_2018}, we use the Yoshida 4th-order symplectic algorithm with an integration step $\Delta t=0.2$ keeping a relative error in the total conserved energy smaller than $\Delta E/E = 10^{-5}$. We choose the so-called 'water bag' initial conditions, which consist of: setting all particle positions $\theta_i$ at zero and uniformly distributing their momenta, simultaneously preserving two constants of motion: the total angular momentum $P=\sum_{i=1}^Np_i=0$ and the desired energy $U=E/N$.  

In order to relate our results to those obtained by other authors, throughout the whole paper, we use $U=0.69$. There are some reasons, which make the HMF model at this energy value an interesting research object. First, in the energy range $U\in(0.5,U_c)$, non-equivalence of ensembles (microcanonical and canonical) is observed, which is most visible for $U=0.69$. (The issues have been intensively studied over the last decade \cite{bookCampa}.) Secondly, in this energy range, QSS was shown to exhibit super-diffusive properties and the findings put forward in the Letter~\cite{1999PRLLatora}, which we intend to refer to, were formulated based on numerical simulations made for this particular energy value. 

\begin{figure}
	\includegraphics[width=\columnwidth]{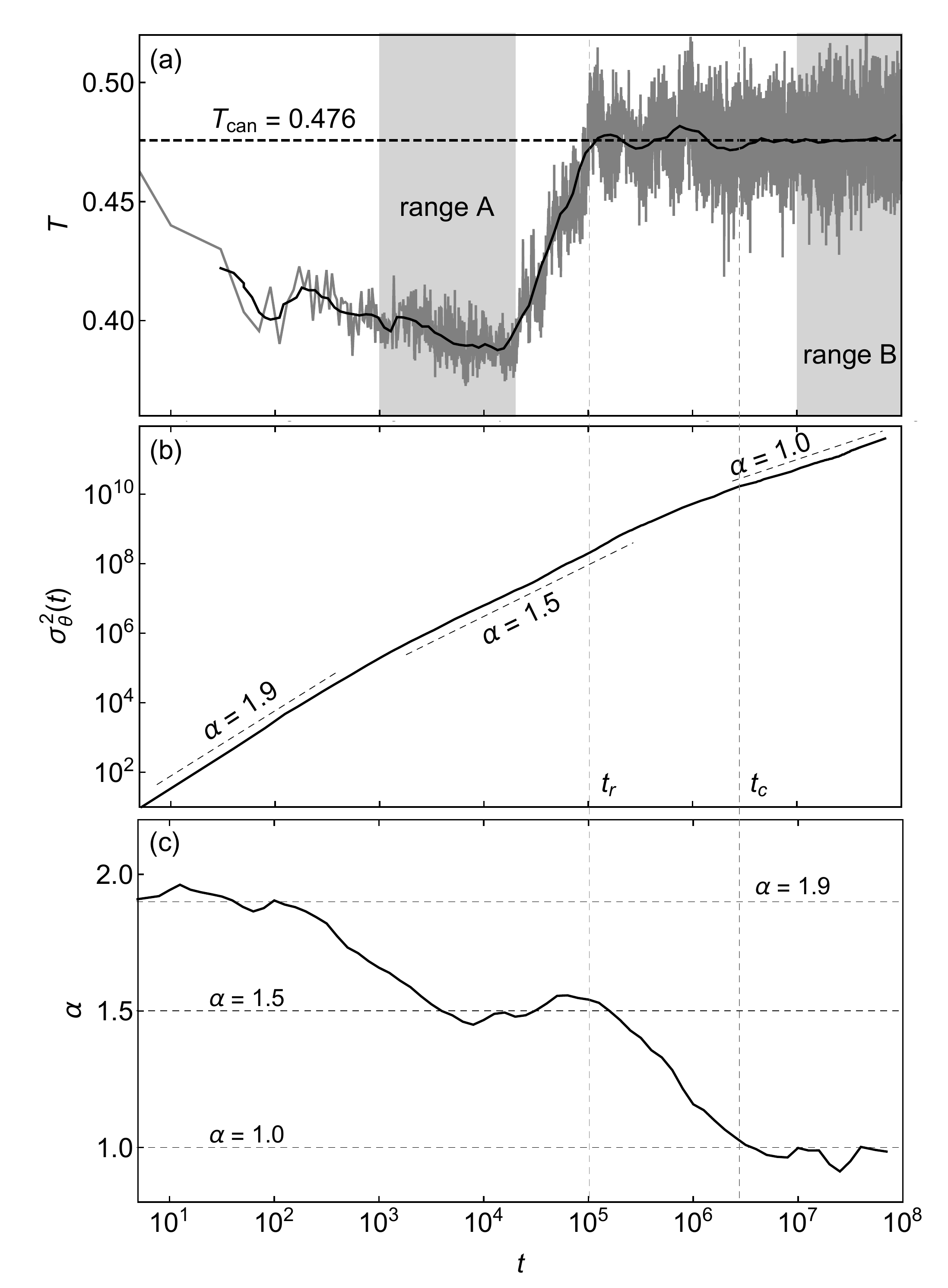}
	\caption{\label{fig1} (a) Temporal behavior of the kinetic temperature. The solid line is the moving average. The dashed line represents the theoretical canonical value of the temperature. The two gray areas indicate regions used for the further statistics - the QSS state and the canonical equilibrium state. (b) Variance of the one-particle angle, $\sigma_\theta^2(t)$. (c) Diffusion exponent $\alpha(t)$ calculated from the data shown in (b).}
\end{figure}

Thus, starting the molecular dynamic simulations of the HMF model from the 'water bag' initial conditions with $U=0.69$, the model evolves toward thermodynamic equilibrium. It is well known, however, that before approaching the BG regime, for some time, it stays in QSS, in which its kinetic temperature, $T=2K/N$, is clearly lower than that of equilibrium, $T_{can}=0.476$. Fig.~\ref{fig1}(a) shows how the temperature changes over time. In this figure, the metastable QSS is marked as 'region A', which, approximately, extends from $t\simeq 10^3$ to $2\times 10^4$. After leaving QSS, the model rapidly changes its kinetic temperature and, finally, after the relaxation time, $t_r\simeq 10^5$, it settles in the stable BG equilibrium. 

In Ref.~\cite{1999PRLLatora}, it is shown that the above-described stages of the HMF model dynamics, which are clearly visible on the temperature vs time plot, are also easy to recognize at the molecular level, where they reveal themselves as super-diffusion (QSS) or normal-diffusion (BG) in the motion of particles. More precisely, the above findings have been confirmed by the time behavior of the average square displacement (see Fig.~\ref{fig1}(b,c)),
\begin{equation}\label{sigmaalfa}
\sigma^2_{\theta}(t)=\langle\theta^2\rangle-\langle\theta\rangle^2\propto t^\alpha,
\end{equation}
which was shown to be super-diffusive (with $\alpha>1$) in QSS, and normal (with $\alpha\simeq1$) in the BG regime. The crossover time, $t_c$, from super-diffusion to normal diffusion, was found to coincide with the relaxation time $t_r$, suggesting the bottom-up causal relation between the two phenomena (i.e., super-diffusion and metastability).

Correspondingly, in Ref.~\cite{1999PRLLatora}, the super-diffusive properties of the HMF model in QSS were recognized as due to L\'evy walks that intertwine, in a particle trajectory, with periods of a quasi-regular oscillating motion. It was shown that these walks disappear when the system approaches equilibrium, which, in turn, was claimed to explain the value of $\alpha\simeq 1$ standing behind the normal diffusion. Moreover, in line with a general model of super-diffusion introduced in \cite{Klafter_1994}, the value of $\alpha$ was suggested to result from the characteristic exponents in power-law distributions describing the duration of 'walk'-like and 'rest'-like periods. In what follows, we present the results of our more extensive numerical simulations of the HMF model that reveal serious misinterpretations of \cite{1999PRLLatora}. First, we show that it is not true that the super-diffusive properties of the HMF model disappear in BG equilibrium. Second, we explain why the parameter $\alpha$ decreases over time to unity, and why, in the considered case, the limiting value of $\alpha=1$ should not be associated with the normal transport. 

Given the results of our numerical simulations, the first clear evidence that the findings of~\cite{1999PRLLatora} are not entirely correct is the difference between the relaxation time, $t_r\simeq 10^5$, and the crossover time, $t_c\simeq 2\times10^6$, observed for the system size $N=2048$ (see Fig. \ref{fig1}(a,b,c)). In Ref.~\cite{1999PRLLatora}, smaller in size and less averaged systems were studied in which the difference, $t_r\neq t_c$, may not have been as clear as in our case. To determine where the differences between our results and those described in \cite{1999PRLLatora} come from, we examined the dynamics of the HMF model at the level of individual particles. Unexpectedly for ourselves, we have discovered that L\'evy walks are present also in the BG regime. 

\begin{figure}
	\includegraphics[width=\columnwidth]{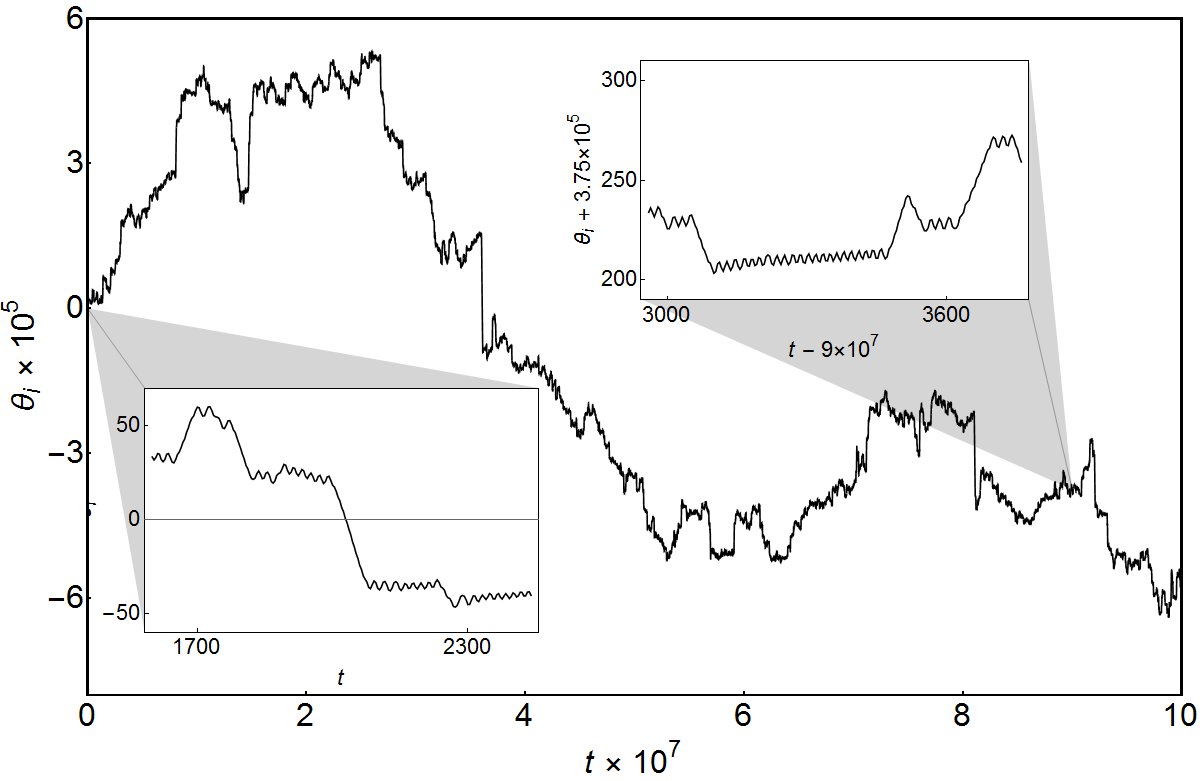}
	\caption{\label{fig2}Typical trajectory of the HMF particle $\theta_i(t)$. Insets magnify trajectories in QSS (left) and the BG regime (right).}
\end{figure}

\begin{figure}
	\includegraphics[width=\columnwidth]{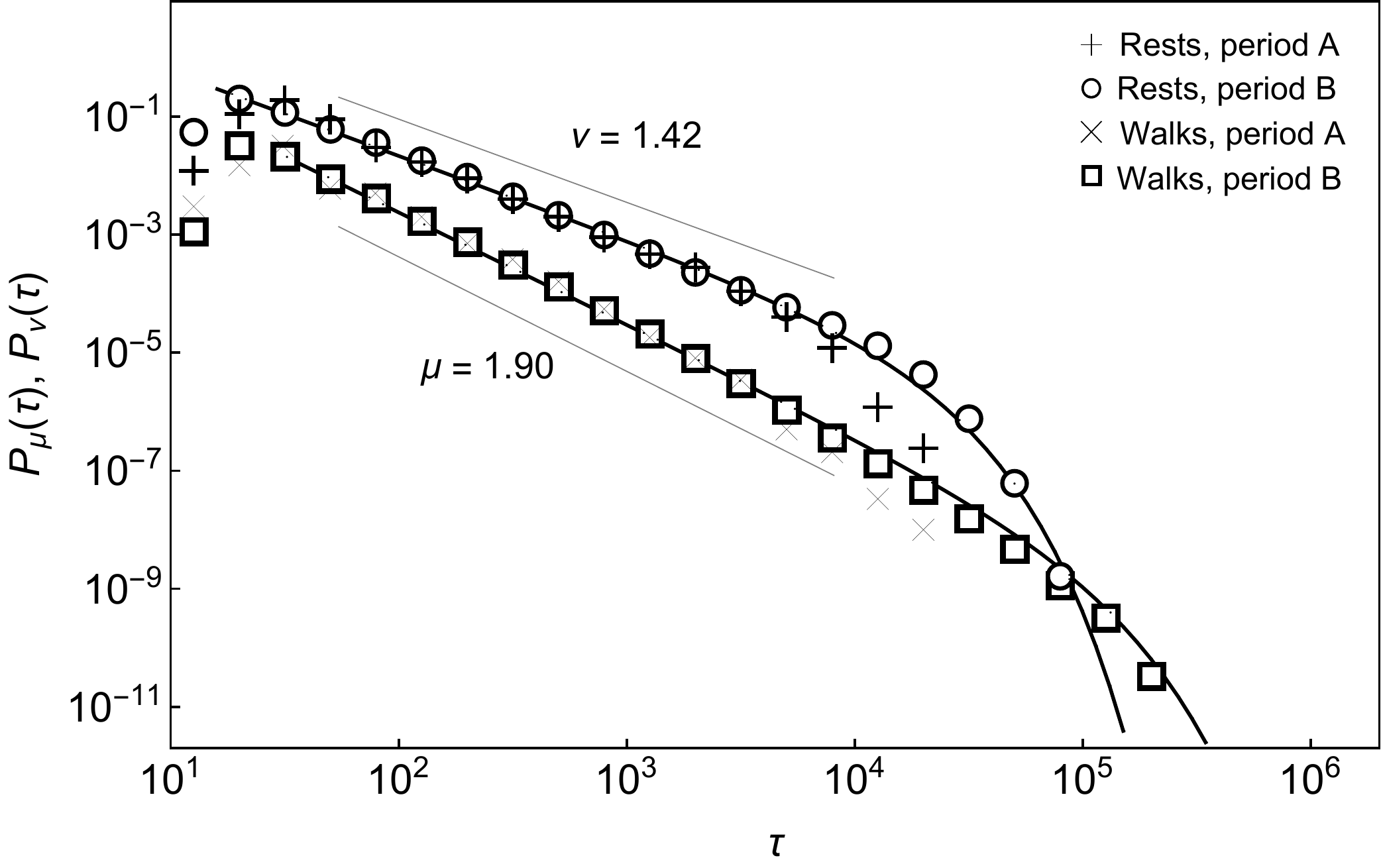}
	\caption{\label{fig3}The probability distributions for trapping and walking times in the BG equilibrium (circles and squares, respectively) and in the QSS ('$+$' and '$\times$' markers, respectively). Solid lines show fittings to the Eq. (\ref{SFexp}). }
\end{figure}

In Fig.~\ref{fig2}, the trajectory, $\theta_i(t)$, of a typical HMF particle is shown (cf. Fig.~3 in \cite{1999PRLLatora}). Interestingly, regardless of whether the system is in QSS or in BG equilibrium, the trajectory looks qualitatively the same (see insets in Fig.~\ref{fig2}). Quantitative analysis of all the trajectories in both stages of the model dynamics confirms the above observation. Specifically, in Fig.~\ref{fig3}, the numerically obtained probability distributions are shown, which characterize, respectively, duration of walking and trapping periods in QSS ('range A' in Fig.~\ref{fig1}(a): $t\in(10^3,2\times 10^4)$) and in the BG equilibrium ('range B' in Fig.~\ref{fig1}(a): $t\in(10^7,10^8)$) - four data sets in total. It is significant that the distributions characterizing the BG equilibrium are almost identical to the distributions that characterize QSS. The data sets shown are well fitted by power-laws with an exponential cut-off:
\begin{equation}
P_{\gamma}(\tau)\sim \tau^{-\gamma}e^{-\lambda \tau},
\label{SFexp}
\end{equation}
where $\gamma\in\{\mu,\nu\}$ and $\lambda\in\{\lambda_\mu,\lambda_\nu\}$ for walking and resting times, respectively. In Fig.~\ref{fig3}, the solid lines represent model fittings to the results obtained from the HMF simulation using parameter estimates: $\mu=1.90$, $\lambda_\mu=0.000015$, $\nu=1.42$ and $\lambda_\nu=0.00008$. 

The new results indicate that the time behavior of parameter $\alpha$ in Eq.~(\ref{sigmaalfa}), and especially its decrease from $\alpha>1$ to $\alpha\simeq1$ is not associated with the transition QSS~$\rightarrow$~BG (cf.~Fig.~\ref{fig1}). In fact, the limiting value of $\alpha=1$  is due to the exponential cut-off in the power-law distribution of L\'evy walks, $P_\mu(\tau)$, Eq.~(\ref{SFexp}). Such a truncated distribution has finite moments and, according to the generalized central limit theorem, after many such walks, positions of the HMF particles will eventually be distributed according to the normal distribution, leading to the observed scaling. The outlined scenario for the time behavior of $\alpha(t)$ is in line with the already classic ideas of Mantegna and Stanley \cite{Mantegna_1994}, who were the first to use truncated L\'evy distributions in describing complex (financial and economic) time series.

To support the above heuristic discussion, we have devised a variant of a stochastic process, called a truncated L\'{e}vy-walk model with rests \cite{Zaburdaev_2014}, which mimics the time behavior of a single HMF particle. In this model, with the probability $q$, the particle walks (i.e. it moves with a constant momentum $\pm p$) for a time $\tau$ drawn from the probability distribution $P_\mu(\tau)$, and, with the probability $1-q$ it rests (i.e., it does not move) for a time drawn from the distribution $P_\nu(\tau)$. If two consecutive moves happen, the direction of motion is altered, $p\rightarrow-p$. All the parameters of the process can be deduced from the molecular dynamics of the HMF model.  

\begin{figure}
	\includegraphics[width=\columnwidth]{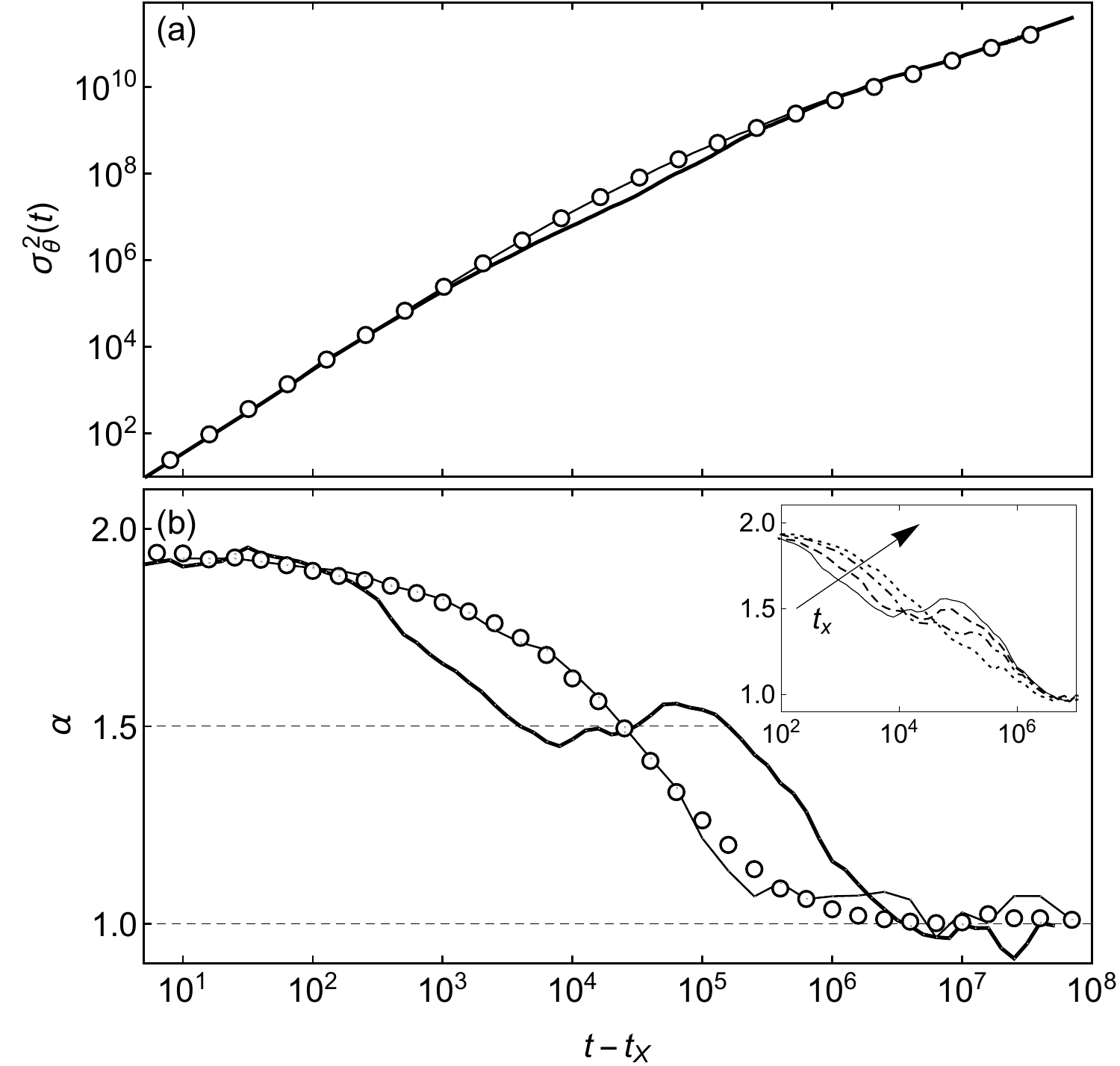}
	\caption{\label{fig4} (a) Time behavior of the mean square displacement, $\sigma_\theta^2(t-t_x)$. The thick curve represents the same data as in Fig.~\ref{fig1}(b) (for $t_x=0$). The thin curve represents the variance after renormalization at $t_x=t_B=10^7$, as described in the text. Open circles represent the values obtained from the similarly renormalized stochastic process and averaged over 100 runs. (b) Diffusion exponent $\alpha(t)$ calculated from the data shown in (a). Line widths and points keep the same meaning as in (a). Inset: $\alpha(t-t_x)$ for different renormalization times $t_x=0,\;10^4,\;5\times10^4$, and $10^5$. The arrow indicates the direction in which $t_x$ grows.}
\end{figure}

In Fig. \ref{fig4}, the diffusive properties of real HMF particles, which are described by the average square displacement, are compared with the corresponding results obtained from our extremely simple model. In this figure, two largely overlapping curves correspond to molecular dynamics simulations of the HMF model, while open circles stand for the stochastic process imitating the motion of HMF particles. Among the curves that describe the real HMF dynamics, the bold curves are exactly the same as shown in Fig.~\ref{fig1}(b,c). The thinner curves relate only to the BG regime ('range B' in Fig. \ref{fig1}(a)), and they were obtained in the following way: At time $t_{B}=10^7$, a kind of renormalization of all particle positions is performed that consists of the modulo division operation: $\theta_i(t-t_B):=\theta_i(t)\mbox{ mod }2\pi$. Since the particle positions are actually angles, such a renormalization procedure has no effect on the dynamics of the system, and it allows to study dispersion of particle trajectories at various stages of the HMF model dynamics. 

With reference to Fig. \ref{fig4}, it is remarkable that the data shown overlap to a large extent. It means that the stochastic process we constructed is sufficient to demonstrate the transition from the super-diffusive regime, $\alpha>1$, to the normal diffusion, $\alpha=1$. In particular, the truncated L\'{e}vy-walk model with rests is well-suited to describe the HMF dynamics in equilibrium. Another point to note here is that the non-stationary dynamics of the HMF model, associated with the transition QSS~$\rightarrow$~BG, leaves an imprint on the time dependence of $\alpha$, which is observed as a plateau $\alpha\simeq 1.5$. We have checked that the width of this plateau depends on the time, $t_x$, when the particle positions are renormalized. The plateau is the widest when $t_x=0$ (which corresponds to non renormalization) and it completely disappears for $t_x>t_r$ (when the system is already in the BG equilibrium). The effect of shortening this plateau is shown in the insets in~Fig. \ref{fig4}. Unfortunately, apart from noticing this shortening, despite many attempts we made (including the analysis of the intensity of various events, correlations between particles, etc.), we have failed to understand the microscopic causes of this plateau. 

In summary, the basic message behind this Letter states that: Equilibrium dynamics of Hamiltonian systems may be super-diffusive. We show that this type of dynamics characterizes the HMF model for the energy value $U=0.69$ (which is slightly smaller than the critical energy at which the second-order phase transition is observed, and for which the nonequivalence of statistical ensembles is most evident). Until now, there has been a strong belief that in the HMF model such dynamics occurs only in QSS. With the help of extensive molecular dynamics simulations, we demonstrate that L\'evy walks, which underlie super-diffusion in QSS, do not disappear after the system settles in BG regime. Furthermore, we show that by only examining the statistics of L\'evy walks in particle trajectories, it is extremely difficult to distinguish QSS from the BG equilibrium. 

The HMF model considered in this paper is an important representative of a broad class of non-additive models, whose physics we are just learning. Extremely slow relaxation processes, non-equivalence of statistical ensembles and negative specific heat are the most spectacular features of these systems. With this paper we add another intriguing feature to this set, which is its super-diffusive dynamics in BG equilibrium states. The L\'evy walks observed therein may partially explain the recent observations made by Cirto et al. \cite{Cirto_2018}, who applied the nonextensive statistical mechanics formalism introduced by Tsallis \cite{Tsallis_2009} to the so-called $\alpha$-HMF model and found that distribution of particles' momenta in the BG regime are $q$-Gaussian rather than Maxwellian, just like in QSS~\cite{Cirto_2014}.

This work has been supported by the National Science
Centre of Poland (Narodowe Centrum Nauki, NCN) under grant no. 2015/18/E/ST2/00560 (A.F. and A. Ch.).

\end{document}